\documentclass[prl,twocolumn,amsmath,amssymb,aps,superscriptaddress,preprintnumbers]{revtex4-2}

\usepackage{graphicx}
\usepackage{braket,slashed,changepage,placeins,multirow,colortbl,makecell,mathtools,array}
\usepackage{xspace}
\usepackage{enumitem}
\usepackage[svgnames]{xcolor}
\usepackage[T1]{fontenc}
\usepackage[utf8]{inputenc}
\usepackage{float}  
\usepackage[colorlinks=true,citecolor=blue,urlcolor=blue,linkcolor=blue,anchorcolor=blue]{hyperref}
\usepackage{bbm}

\newcommand{\chibar}{\bar{\chi}}

\newcommand{\cC}{\ensuremath{\mathcal{C}}}

\newcommand{\LQCD}{\ensuremath{\Lambda_{\rm QCD}}}

\newcommand{\nbar}{\bar{n}}

\newcommand{\cO}{\ensuremath{\mathcal{O}}}

\newcommand{\psibar}{\bar{\psi}}

\newcommand{\cP}{\ensuremath{\mathcal{P}}}

\newcommand{\Tbar}{\bar{T}}

\newcommand{\xbar}{\bar{x}}

\DeclareMathOperator*{\sumint}{%
	\mathchoice%
	{\ooalign{$\displaystyle\sum$\cr\hidewidth$\displaystyle\int$\hidewidth\cr}}
	{\ooalign{\raisebox{.14\height}{\scalebox{.7}{$\textstyle\sum$}}\cr\hidewidth$\textstyle\int$\hidewidth\cr}}
	{\ooalign{\raisebox{.2\height}{\scalebox{.6}{$\scriptstyle\sum$}}\cr$\scriptstyle\int$\cr}}
	{\ooalign{\raisebox{.2\height}{\scalebox{.6}{$\scriptstyle\sum$}}\cr$\scriptstyle\int$\cr}}
}

\def\nn{{\nonumber}}

\newcommand{\eq}[1]{Eq.~\eqref{eq:#1}}
\newcommand{\eqs}[2]{Eqs.~\eqref{eq:#1} and \eqref{eq:#2}}

\usepackage{lipsum}

\usepackage{marginnote}

\usepackage{xcolor}
\usepackage[normalem]{ulem}
\begin{document}
	
\preprint{MIT-CTP 5948, LA-UR-26-21164}

\title{Minding the gap between hard and diffractive structure functions at the EIC}

\author{June-Haak Ee}
\email{jhee@mit.edu}
\affiliation{MIT Center For Theoretical Physics -- A Leinweber Institute, Cambridge, MA 02139, USA}
\affiliation{Theoretical Division, Los Alamos National Laboratory, Los Alamos, NM 87545, USA}

\author{Christopher Lee}
\email{clee@lanl.gov}
\affiliation{Theoretical Division, Los Alamos National Laboratory, Los Alamos, NM 87545, USA}
\affiliation{Physics Division, Los Alamos National Laboratory, Los Alamos, NM 87545, USA}

\author{Stella T. Schindler}
\email{schindler@lanl.gov}
\affiliation{Theoretical Division, Los Alamos National Laboratory, Los Alamos, NM 87545, USA}
\affiliation{Physics Division, Los Alamos National Laboratory, Los Alamos, NM 87545, USA}

\begin{abstract}
Diffraction is characterized by forward scattering and a large rapidity gap.
Ordinary hard scattering also produces gapped events and presents a significant background to identification of diffractively produced events. We present a fully differential factorization of gapped hard $ep$ scattering. 
We find that the unpolarized azimuthally-dependent structure functions $F_{3,4}^{\rm gap}$ and the longitudinal $F_L^{\rm gap}$ vanish at leading power for hard scattering.
Azimuthal dependence is thus a pure diffractive signature. This is borne out in Monte Carlo simulations. 
\end{abstract}

\maketitle
Twenty percent of collisions at the upcoming Electron-Ion Collider (EIC) at Brookhaven are expected to exhibit a strange feature: a large angular wedge of the detector (a \textit{rapidity gap}) will remain untouched by the particle spray \cite{AbdulKhalek:2021gbh}. Rapidity-gapped events that involve a forward-scattered initial-state hadron are called \textit{diffractive} \cite{Feinberg1956, Chew:1961ev, Frankfurt:2022jns}. Diffraction is among the cleanest probes of gluon saturation \cite{Glauber:1955qq, Gribov:1968jf}, and illuminates physics ranging from forward (Regge) dynamics \cite{Hentschinski:2022xnd} to cosmic-ray cascades \cite{Engel:2011zzb, LHCForwardPhysicsWorkingGroup:2016ote} and beyond-the-Standard-Model dynamics \cite{Khoze:2001xm, Albrow:2010yb, Feng:2022inv, CMS:2023roj}.

A rapidity gap, however, does not by itself signal diffraction. Large gaps also arise in ordinary hard scattering \cite{Kang:2013nha,Dasgupta:2003iq, Kang:2012zr, Berger:2010xi, Banfi:2012yh,Becher:2012qa,Tackmann:2012bt,Banfi:2012jm,Liu:2012sz,Jouttenus:2013hs, Stewart:2013faa,Shao:2013uba,Li:2014ria,Forshaw:2006fk,Becher:2023mtx, Qiu:2022pla, Qiu:2024pmh,Ee:2025scz}, e.g., when a pencil-like two-jet configuration leaves the central region empty. This mechanism is distinct from the forward exchange of diffraction. Because the gap itself often has served as the experimental criterion for selecting diffractive events \cite{H1:2006zyl} (particularly for \textit{incoherent} diffraction, which produces a forward jet), this hard component is an irreducible background that can bias diffractive measurements, and hence the physics extracted from them. The two are hardest to tell apart precisely where the contamination matters most: in large-gap events.

\begin{figure}[t]
    \centering
    \includegraphics[width=3.3in]{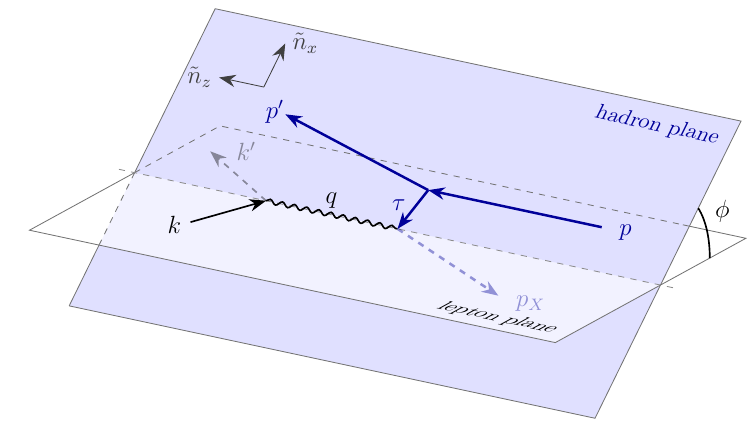}
    \caption{{\bf Breit frame configuration of gapped hard scattering $ep \to e'X_CX_T$, with $X_i$ a collection of one or more particles}. The lepton plane (with $k$, $k'$, $q$) and hadron plane (with $p$, $p'$, $p_X$, $\tau$) intersect along $q$ and are separated by azimuthal angle $\phi$. The longitudinal ($\tilde n_z$) and transverse ($\tilde n_x$, in hadron plane) axes are indicated.}\label{fig:gaps}
    \vspace{-1em} 
\end{figure}

In this paper, we provide a fully differential \textit{factorization formula} for hard $ep$ scattering with two gapped rapidity sectors (Fig.~\ref{fig:gaps}), expressing its cross section in terms of first-principles field theory operators. A typical gap choice is  $\Delta\eta \gtrsim 2$--$3$ \cite{ATLAS:2012djz,ATLAS:2019asg}.
The formula predicts both the rate of gapped hard events and their contamination of diffractive observables. 
Our central result is that the structure functions $F_L^{\rm gap}$, $F_3^{\rm gap}$, and $F_4^{\rm gap}$ vanish at leading power in gapped hard scattering: the hard tensor is azimuthally trivial, so the azimuthal structure functions $F_{3,4}^{\rm gap}$---and the longitudinal $F_L^{\rm gap}$---are pure diffractive signatures in the presence of a gap. Measuring azimuthal dependence therefore cleanly separates genuine diffraction from its hard background.
We perform Monte-Carlo simulations with RAPGAP \cite{Jung:1993gf} which show that the hard $F_{3,4}^{\rm gap}$ switch off in the large-gap limit, leaving azimuthal dependence as a clean diffractive signature.

\noindent \textbf{\textit{Kinematics.}} 
Consider the gapped process $ep \to e'X_CX_T$ in Fig.~\ref{fig:gaps}, where $X_C$ and $X_T$ are sprays of one or more particles (e.g. a jet) in the central and target regions, separated by a rapidity gap. The initial and final electrons carry momenta $k$ and $k'$, while the initial proton and final hadronic state $X_T$ carry momenta $p$ and $p'$, respectively. The lepton and hadron transfer momenta $q = k-k'$ and $\tau = p-p'$ to the central system $X_C$, which carries momentum $p_X = q+\tau$.

We work with vanishing hadronic and leptonic mass
$k^2 = k^{\prime 2} = m_e^2 \approx 0$ and $p^2 = m_p^2 \approx 0$ giving seven independent Lorentz invariants:
\begin{align}
\label{eq:invariants}
	& x = \frac{Q^2}{2p\cdot q}\,,
	&& y = \frac{p\cdot q}{p\cdot k}\,,
	&& Q^2 = -q^2\,,
	\\
	& \beta = \frac{Q^2}{2q\cdot \tau}\,,
	&& t = \tau^2\,,
	&& \xbar = \frac{k\cdot \tau}{k\cdot p}\,,
	& m_T^2 = (p')^2 \,.
	\nn
\end{align}
It is often useful to consider additional invariants that are not independent of the above:
\begin{align}
\label{eq:further-invariants}
	& W^2 = (p+q)^2 = Q^2\frac{1-x}{x}\,,
	&&s = (p+k)^2  = \frac{Q^2}{xy}\,,
	\\
	& m_X^2 = p_X^2 = Q^2\frac{1-\beta}{\beta} + t\,,
	&& z = \frac{p\cdot p'}{p\cdot q} = \frac{x}{Q^2}(m_T^2-t)\,.
	\nn
\end{align}
Ref.~\cite{Lee:2025fml} provides bounds for these invariants.

\noindent \textbf{\textit{Breit frame.}}
We work in the Breit frame, defined as 
\begin{align}
\label{eq:p-q-in-Breit}
	&p^\mu = \frac{Q}{x} \frac{n^\mu}{2}\,,
	&& q^\mu = Q\frac{\nbar^\mu - n^\mu}{2}\,,
\end{align}
up to mass corrections $\sim\cO(m_p^2/Q^2)$. 
In this frame, the light-like vectors $n$ and $\bar{n}$ are $n^\mu = (2/Q)xp^\mu$ and $\bar{n}^\mu = (2/Q)(xp^\mu + q^\mu)$. Then, defining the light-cone coordinates $p^\mu= (n\cdot p, \nbar\cdot p, p_\perp)$, we have $n^\mu = (0,2,0_\perp)$ and $\nbar^\mu = (2,0,0_\perp)$, and 
\begin{align}
\label{eq:tau-px-pp}
	&\tau = Q\Big(-z, \frac{1}{\beta}-z,-\sqrt{z^2 -z/\beta - t/Q^2} \Big)\,,
	\nn\\
	&p_X = Q\Big(1-z, \frac{1-\beta}{\beta}-z,-\sqrt{z^2 -z/\beta - t/Q^2} \Big)\,,
	\nn\\
	&p' = Q \Big( z, \frac{1}{x}-\frac{1}{\beta}+z, \sqrt{z^2  -z/\beta - t/Q^2}\Big).
\end{align}
The transverse axis is taken in the hadron plane  (Fig.~\ref{fig:gaps}), with $p'$ at positive $\perp$ and the recoil $\tau$, $p_X$ at negative $\perp$.
\\[-8 pt]

\noindent \textbf{\textit{Structure functions.}} 
Following \cite{Lee:2025fml}, the fully differential cross sections for diffraction and gapped hard scattering both take the form
\begin{align}
\label{eq:cross-section}
	\frac{d^6\sigma}{dx\,dQ^2\,d\beta\,dt\,dm_T^2\,d\phi} \!=\! \frac{\pi \alpha^2 xy^2}{4Q^6\beta^2} L_{\mu\nu}(k,k') W_{\rm gap}^{\mu\nu}(q,p,p')\,,
\end{align}
where $\phi$ is the relative azimuthal angle of $p'$ with respect to the lepton plane.
The unpolarized leptonic and hadronic tensors are
\begin{align}\label{eq:hadron-lepton}
	&L^{\mu\nu}= 2(k^\mu k^{\prime \nu} + k^\nu k^{\prime\mu} - k\cdot k' \, g^{\mu\nu})\,,
	\nn\\
	&W_{\rm gap}^{\mu\nu} = \sumint_{X_C,X_T} \delta^4(q+p-p'-p_X)\delta(m_T^2 - p^{\prime 2})
	\nn\\
	&\hspace{0.5 cm}\qquad \times \langle p| J^{\dagger \mu}(0)|X_CX_T \rangle \langle X_C X_T| J^\nu(0)|p\rangle \,,
\end{align}
with the vector current $J^\mu = \sum_f e_f\,\psibar_f\gamma^\mu\psi_f$ for quark flavors $f$ of charge $e_f$, and each phase-space sum and integral $\sumint$ running over all final-state configurations, with $\sumint_{X_T}$ normalized to 1 for the one-particle $X_T$ state. We work throughout in the unpolarized, vector--vector case:
electromagnetic currents dominate in the moderate $x=Q^2/sy$ kinematics at the EIC ($\sqrt{s}=20$--$140$~GeV) and HERA ($\sqrt{s} = 319$ GeV) on which we focus. 

As in \cite{Arens:1996xw, Lee:2025fml}, we decompose $W_{\rm gap}^{\mu\nu}$ into structure functions $F_i^{\rm gap}$,
\begin{align}\label{eq:structure-functions}
	W_{\rm gap}^{\mu\nu} = \sum_i w_i^{\mu\nu}(U,X,Z)\, F_i^{\rm gap}(x,Q^2,\beta,t,m_T^2)\,,
\end{align}
with $i\in \{L, 2, 3, 4\}$.
These $F_i^{\rm gap}$ have dependence on $p'$ through the invariants in \eq{invariants}, unlike the structure functions $F_i$ of DIS.
Here $U^\mu$, $X^\mu$, $Z^\mu$ are orthonormal vectors:
\begin{align}
\label{eq:UXZ-def}
	U^{\mu} &= \frac{2x}{Q}\left(p^\mu + \frac{q^\mu }{2x} \right)\,,
	\nonumber\\
	X^\mu &= \frac{(\frac{1}{\beta}-2 z)xp^\mu - zq^\mu - \tau^\mu}{Q\sqrt{z^2 -z/\beta - t/Q^2}}\,,
    \quad
    Z^\mu = -\frac{q^\mu}{Q},
\end{align}
which satisfy $U^2=1$, $X^2=Z^2=-1$ and $U\cdot X=U\cdot Z=Z\cdot X=0$. 
In the Breit frame, $X^\mu = \tilde n_x^\mu=(0,1,0,0)$ and $Z^\mu = \tilde n_z^\mu=(0,0,0,1)$ are the transverse and longitudinal spatial axes shown in Fig.~\ref{fig:gaps}, while $U^\mu = \tilde n_t^\mu=(1,0,0,0)$ is the time direction.
Defining the remaining spacelike direction normal to the hadron plane (not shown in Fig.~\ref{fig:gaps}),
\begin{align}
	Y^\mu = \epsilon^{\mu\nu\rho\sigma} U_\nu X_\rho Z_\sigma\,,
	\qquad Y^2 = -1\,,
\end{align}
which in the Breit frame is $Y^\mu = \tilde n_y^\mu = (0,0,1,0)$, the four vectors form a complete orthonormal basis (with $U$ timelike) valid in any frame:
\begin{align}\label{eq:completeness}
	g^{\mu\nu} = U^\mu U^\nu - X^\mu X^\nu - Y^\mu Y^\nu - Z^\mu Z^\nu\,.
\end{align}
The tensors $w_i^{\mu\nu}$ are then constructed from these basis vectors as 
\begin{align}
	w_L^{\mu\nu} &= 1/(2x)(U^\mu U^\nu - X^\mu X^\nu - Y^\mu Y^\nu)\,,
    \nn\\
	w_2^{\mu\nu} &= 1/(2x)(X^\mu X^\nu + Y^\mu Y^\nu)\,,
	\nn\\
    w_3^{\mu\nu} &= 1/(2x)(X^\mu X^\nu- Y^\mu Y^\nu)\,,
	\nn\\
	w_4^{\mu\nu} &= 1/(2x)(U^\mu X^\nu + X^\mu U^\nu)\,, 
\end{align}
where we use the completeness relation in Eq.~\eqref{eq:completeness}.
These have projectors $\cP_i$ satisfying $\cP_{i\mu\nu}\,w_j^{\mu\nu}=\delta_{ij}$ (see Ref.~\cite{Lee:2025fml}), 
so that $\cP_{i\mu\nu}W_{\rm gap}^{\mu\nu}=F_i^{\rm gap}$.
The azimuthal functions $F_{3,4}^{\rm gap}$ probe the $X$--$Y$
anisotropy of $W_{\rm gap}^{\mu\nu}$, which we show below vanish at leading power in hard scattering.
Note that the diffractive naming convention \cite{Lee:2025fml} for these parity-conserving $F_{3,4}^{\rm gap}$ is distinct from the traditional parity-violating structure function $F_{3}$ of DIS.

In the Breit frame, the azimuthal angle $\phi$ of $p'_\perp$ about $q$, measured from the lepton plane (Fig.~\ref{fig:gaps}), is fixed by
\begin{align}
k\cdot X&=k\cdot \tilde n_x = -|k_\perp| \cos\phi,
\nonumber \\
k\cdot Y&=k\cdot \tilde n_y = -|k_\perp| \sin\phi,
\end{align}
where the minus signs arise from the spacelike metric.  
Then, from $|k_\perp|=Q\sqrt{1-y}/y$ in the Breit frame and the explicit expression for $X^\mu$ in Eq.~\eqref{eq:UXZ-def}, we can write $\cos\phi$ in terms of Lorentz invariants as
\begin{equation}
\label{eq:cosphi-xbar}
\cos\phi = 
\frac{\bar{x} - x/\beta+ (2-y)xz}
{2x\sqrt{1-y}\sqrt{z^2-z/\beta-t/Q^2}}.
\end{equation}
The leptonic--hadronic contraction then becomes
\begin{align}
\label{eq:coefficients}
	L_{\mu\nu}W_{\rm gap}^{\mu\nu} 
    &= 
    \frac{2s}{y} \big[-\frac{y^2}{2} F_L^{\rm gap} + \big(1-y+ \frac{y^2}{2}\big)F_2^{\rm gap} 
	\\
	&\hspace{-1 cm}+ (1-y)\cos (2\phi) F_3^{\rm gap} - (2-y)\sqrt{1-y} \cos\phi F_4^{\rm gap}\big]\,. 
	\nn
\end{align}
For diffractive kinematics, $F_2^{\rm gap}$ was measured often at HERA; $F_L^{\rm gap}$ less so \cite{H1:2011jpo}. Measuring $F_{3,4}^{\rm gap}$ would require sufficient resolution in $\xbar$; see \S2.4 of~\cite{Lee:2025fml} and \cite{ZEUS:2004luu}.
\\[-8 pt]

\begin{table}[t!]
	\begin{center}\renewcommand{\arraystretch}{1.2}
		\begin{tabular}{|c|c|c|}
			\hline
			\rowcolor{WhiteSmoke} {\bf Invariant} &{\bf Gapped hard} & {\bf Diffraction }    
			\\ \hline
			$x$&$x\sim \mathcal{O}(1)$ & $x \ll 1$
			\\ 
			$\beta$ & $1-\beta \ll 1$ & $\beta \gg x$, with $\beta \sim 1$ or $\beta \ll 1$
			\\ 
			$m_X, m_T$ & $m_X^2, m_T^2\ll Q^2$ & $m_X^2, m_T^2 \ll W^2$
			\\
			$t$ & $-t\ll Q^2 \sim s$ & $-t \ll W^2$
			\\ \hline
		\end{tabular}
	\end{center}
	\caption{Behavior of 
    Lorentz invariants, defined in \eqs{invariants}{further-invariants}, in gapped hard scattering vs. diffraction with two measured final-state regions separated by a rapidity gap 
    \cite{Lee:2025fml}.}\label{tab:gaps}
\end{table}

\noindent \textbf{\textit{Kinematic constraints.}}  
We now quantify what rapidity-gapped hard scattering requires kinematically. First,
the $ep$ collision must be hard; second, $p'$ and $p_X$ must form two distinct,
low-mass hadronic sectors; third, there must be a rapidity gap between them, with
$\eta(k)=\tfrac12\ln(k^-/k^+)$, $k^+=n\cdot k$, $k^-=\bar n\cdot k$. These give
\begin{align}\label{eq:constraints}
	&x\sim 1\,,
	&&\LQCD^2 \ll m_X^2, m_T^2 \ll Q^2\,,
	&&\frac{p_X^-}{p_X^+} \ll \frac{p^{\prime -}}{p^{\prime +}} \,.
\end{align}
Their consequences for the invariants are immediate: combining $x\sim1$ with
$m_X^2\ll Q^2$ gives $1-\beta\ll1$, whence $-t = Q^2(1-\beta)/\beta - m_X^2 \ll Q^2$.
The full set of conditions for gapped hard scattering and diffraction~\cite{Lee:2025fml}
is collected in Table~\ref{tab:gaps}.

We encode these scalings in \textit{power-counting} parameters
\begin{align}
	& \lambda = \frac{Q}{\sqrt s}\,,\;\;
	 \lambda_t = \frac{\sqrt{-t}}{Q}\,,\;\;
	 \rho = \frac{m_T}{\sqrt{-t}}\,,\;\;
	 \lambda_\Lambda = \frac{\LQCD}{\sqrt{-t}}\,,
\end{align}
in terms of which the constraints read
\begin{align}\label{eq:constraints-pc}
	&\lambda\sim1\,,
	&& \lambda_t\ll1\,,
	&& \lambda_t\rho\ll1\,,
	&& \lambda_\Lambda\ll1\,,
\end{align}
together with $1-\beta\ll1$. We can then express the Breit-frame momenta in terms of their scaling in $\lambda_t$
\begin{align}\label{eq:modes}
	& p \sim \sqrt s\,(0,1,0)\,,
	&& q \sim \sqrt s\,(1,1,0)\,,
	\nn\\
	& p_X \sim \sqrt s\,(1,\lambda_t^2,\lambda_t)\,,
	&& \tau \sim \sqrt s\,(\lambda_t^2,1,\lambda_t)\,,
	\nn\\
	& p' \sim \sqrt s\,(\lambda_t^2,1,\lambda_t)\,.
\end{align}
Here, we drop $\cO(\lambda_\Lambda)$ terms and consider $1-\beta\sim\lambda_t^2$, $\rho\sim1$
for simplicity: Eq.~\eqref{eq:constraints-pc} indicates that analyzing the hard interaction will require us to expand in $\lambda_t\ll1$, but does not require such an expansion for the other parameters.

Equation~\eqref{eq:modes} enables us to \textit{factorize} $W_{\rm gap}^{\mu\nu}$, i.e., to separate the physics of different momentum scales into field theory matrix elements with clear physical interpretation.
We use the technique of \textit{effective field theory} (EFT), which systematically power-expands the QCD Lagrangian in $\lambda_t\ll1$ about relevant modes.  
The modes in Eq.~\eqref{eq:modes} belong to the
soft-collinear effective theory
(SCET)~\cite{Bauer:2000ew, Bauer:2000yr, Bauer:2001ct, Bauer:2001yt, Rothstein:2016bsq}:
$p$, $\tau$, $p'$ are $n$-collinear, $p_X$ is $\nbar$-collinear, and the exchanged photon
$q$ is hard. Note that these modes can also radiate ultrasoft modes of momentum scaling $\sim \sqrt{s} (\lambda_t^2, \lambda_t^2,\lambda_t^2)$ without changing the invariant mass scaling of $X_{T,C}$, whereas soft modes $\sim\sqrt{s}(\lambda_t,\lambda_t,\lambda_t)$ would; the mass measurements thus select ultrasoft radiation, which we must build into our formalism.

\noindent \textbf{\textit{Factorization.}}
The factorization of $W_{\rm gap}^{\mu\nu}$ in Eq.~\eqref{eq:hadron-lepton} closely parallels the fully
differential form of the $1$-jettiness factorization of Ref.~\cite{Kang:2013nha}, which
likewise describes hard scattering with two rapidity-separated sectors. We sketch its structure here, and provide a complete derivation in the Supplemental Material~\cite{SupplementalMaterial}\nocite{Becher:2006qw,Jain:2011iu,Kelley:2011ng,Monni:2011gb,Hornig:2011iu,Manohar:2003vb,Idilbi:2006dg,Becher:2006mr}.

Using SCET, the states in Eq.~\eqref{eq:hadron-lepton} factorize into $n$-collinear,
$\nbar$-collinear, and ultrasoft ($us$) components:
\begin{align}\label{eq:factorize-states}
	&\langle p | \!=\! \langle p_n | \langle 0_{\nbar} | \langle 0_{us}|,
	&&| X_CX_T \rangle \!=\! |0_{us}\rangle | X_{C,\nbar} \rangle | X_{T,n} \rangle.
\end{align}
The electromagnetic current matches onto a hard coefficient $\cC^\nu$ connecting the
$n$-collinear (beam) and $\nbar$-collinear (current) sectors,
$J^\nu \to \sum_f e_f\,\chibar_{\nbar} \cC^\nu \chi_n$, with $\chi_n$ the $n$-collinear
quark field in SCET. Decoupling collinear and ultrasoft fields by the standard BPS field
redefinition~\cite{Bauer:2001yt}, $\chi_n \to Y_n \chi_n$ with $Y_n$ an ultrasoft gluon
Wilson line, gives
\begin{align}\label{eq:factorize-currents}
	J^\nu = \sum_f e_f \, \cC^\nu \chibar_{\nbar} \,T[Y_{\nbar}^\dagger Y_n]\, \chi_n \,.
\end{align}
Inserting Eqs.~\eqref{eq:factorize-states} and \eqref{eq:factorize-currents} into
Eq.~\eqref{eq:hadron-lepton}, we have schematically that, up to subleading corrections in $\lambda_t$,
\begin{align}
	&W_{\rm gap}^{\mu\nu}(q,p,p') =  \cC^\nu \cC^\mu  \sumint_{X_C} \langle 0_{\nbar} | \chi_{\nbar}| X_{C,\nbar} \rangle  \langle X_{C,\nbar} |  \chibar_{\nbar}| 0_{\nbar} \rangle
	\nn\\
	& \times\sumint_{X_T} \langle p_n | \chibar_n | X_{T,n} \rangle \langle X_{T,n} |  \chi_n | p_n \rangle \delta(m_T^2 - p^{\prime 2})
	\nn\\
	& \times
    \langle 0_{us}|  \Tbar[Y_n^\dagger Y_{\nbar}]\, T[Y_{\nbar}^\dagger Y_n] |0_{us}\rangle
	\delta^4(q+p-p'-p_X),
\end{align}
where we have pre-emptively grouped terms of the same scaling together and collected
$\sum_f e_f^2$ into the hard coefficients. This simplifies to
\begin{align}\label{eq:factorization}
	W_{\rm gap}^{\mu\nu} = H^{\mu\nu} \otimes \mathcal{B} \otimes J \otimes S\,,
\end{align}
where the hard ($H$), beam ($\mathcal{B}$), jet ($J$), and soft functions are, schematically,
\begin{align}
	&H^{\mu\nu} = \cC^\nu \cC^\mu\,,
	&& \mathcal{B} = \langle p_n | \chibar_n \chi_n | p_n \rangle \,,
	\nn\\
	&J = \langle 0| \chi_{\nbar} \chibar_{\nbar} | 0 \rangle \,,
	&& S = \langle 0 | \Tbar[Y_n^\dagger Y_{\nbar}]\, T[Y_{\nbar}^\dagger Y_n] | 0 \rangle \,.
\end{align}
The gap measurements enter as delta functions fixing the sector virtualities: the jet
virtuality follows internally as $m_X^2=Q^2(1-\beta)/\beta+t$, entering the jet function
$J$, while $\mathcal{B}$ coincides with the standard $1$-jettiness fully-differential beam function, with its arguments fixed by the
$m_T^2$ and $t$ measurement delta functions. The shared soft momenta thread $S$,
assembling the fully differential cross section. [See the Supplemental Material~\cite{SupplementalMaterial} for the full expression.]
Imposing genuinely new constraints---e.g.\ a further
rapidity gap within a sector, taking an exclusive final state, or a different $\rho$ scaling---can refactorize $J$ or $\mathcal{B}$ into additional
matrix elements. All Lorentz indices reside in $H^{\mu\nu}$, which we turn to next.
\\[-8pt]

\noindent \textbf{\textit{Vanishing structure functions.}}
Since $H^{\mu\nu}$ carries all the Lorentz indices of Eq.~\eqref{eq:factorization},
the structure-function content is fixed by its tensor structure alone. At leading
power, for the vector--vector currents of Eq.~\eqref{eq:hadron-lepton},
Ref.~\cite{Kang:2013nha} gives $H^{\mu\nu}= -g_\perp^{\mu\nu}\,H(Q^2,\mu)$ with a scalar
coefficient $H(Q^2,\mu)$, where $g_\perp^{\mu\nu}$ is the metric transverse to the beam
and jet collinear directions $n_B$, $n_J$ of the SCET operator,
\begin{align}
\label{eq:hard-tensor}
  g_\perp^{\mu\nu} = g^{\mu\nu} - \frac{n_B^\mu n_J^\nu + n_J^\mu n_B^\nu}{n_B\cdot n_J}\,,
\end{align}
satisfying $g_\perp^{\mu\nu}n_{B\nu}=g_\perp^{\mu\nu}n_{J\nu}=0$ by virtue of
$n_B^2=n_J^2=0$. At leading power $n_B\parallel n$ and $n_J\parallel\nbar$ [the
directions of $q_B$, $q_J$], with $n\cdot\nbar=2$. The longitudinal pair spans the same
plane as $\{n,\nbar\}$: explicitly, $U^\mu=\tfrac12(n^\mu+\nbar^\mu)$ and
$Z^\mu=\tfrac12(n^\mu-\nbar^\mu)$, so
$\tfrac12(n^\mu\nbar^\nu+\nbar^\mu n^\nu)=U^\mu U^\nu-Z^\mu Z^\nu$ and the completeness
relation~\eqref{eq:completeness} collapses the transverse metric to
\begin{align}
  \label{eq:gperp-XY}
  g_\perp^{\mu\nu} = -(X^\mu X^\nu + Y^\mu Y^\nu) = -2x\, w_2^{\mu\nu}\,.
\end{align}
The leading hard tensor is thus purely transverse and isotropic in the $X$--$Y$ plane,
proportional to $w_2^{\mu\nu}$ alone. Applying the projectors $\cP_i$ then feeds it into $F_2^{\rm gap}$ alone:
\begin{align}
  \label{eq:vanishing}
  F_L^{\rm gap} = F_3^{\rm gap} = F_4^{\rm gap} = 0
\end{align}
exactly at leading power. The first nonzero $F_{L,3,4}^{\rm gap}$ are suppressed by powers of $\lambda_t$ relative to $F_2^{\rm gap}$. The vanishing of $F_{3,4}^{\rm gap}$ is manifest in Eq.~\eqref{eq:gperp-XY}: $g_\perp^{\mu\nu}$ contains
neither the $X$--$Y$ anisotropy ($X^\mu X^\nu-Y^\mu Y^\nu$) nor the $U$--$X$ mixing
($U^\mu X^\nu+X^\mu U^\nu$) that define $w_3^{\mu\nu}$ and $w_4^{\mu\nu}$.
\\[-8pt]

\begin{figure*}[t]
  \centering
  \includegraphics[width=0.87\linewidth]{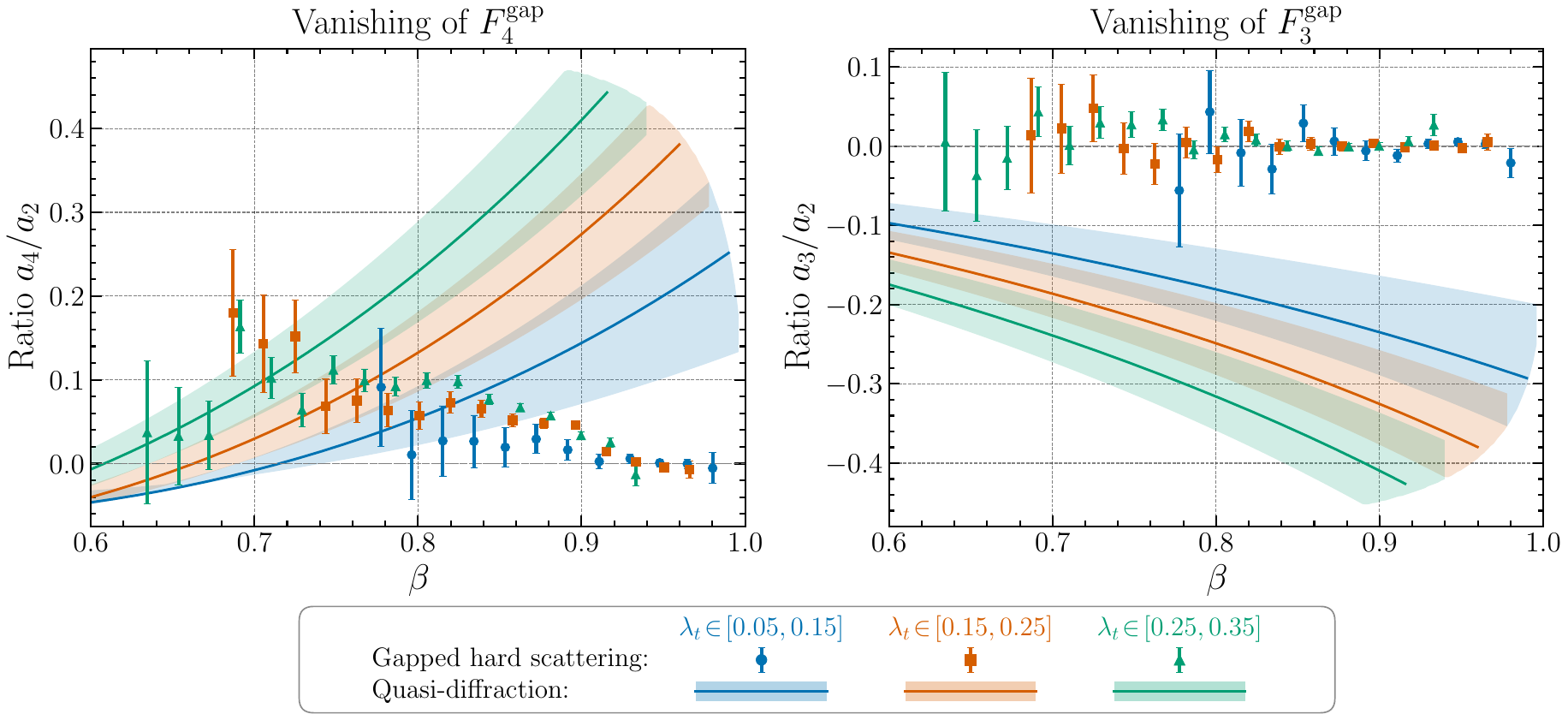}
  \caption{{\bf Ratios involving $F_4^{\rm gap}$\!\! (left) and $F_3^{\rm gap}$\!\! (right), for gapped hard scattering (points) and quasi-diffraction (bands).} Specifically, we plot the ratios $a_{3,4}/a_2 \propto F_{3,4}^{\rm gap}/(c_2 F_2^{\rm gap} + c_L F_L^{\rm gap})$ in Eq.~\eqref{eq:fit}, where $c_{2/L}$ indicates the $y$-dependent coefficients in Eq.~\eqref{eq:coefficients}.
  Ratios are plotted against $\beta$ in narrow windows of $x\in(0.02,0.04)$ and
  $Q^2\in(800,1000)~\mathrm{GeV}^2$. 
  Gapped hard scattering data are obtained from RAPGAP, as described in the text; they are horizontally offset within each $\beta$ bin for visibility. They give a very different shape for the distributions than quasi-diffraction.
  Because each $\lambda_t$ slice is kinematically confined to a limited $\beta$ range (small $\lambda_t$ favoring large $\beta$), bins with insufficient statistics for a reliable $\phi$ fit are not shown.
  The bands span the leading-power quasi-diffractive prediction of Ref.~\cite{Lee:2025fml} over the $\lambda_t$ range of each slice and the $x$ window
  (at $Q^2=900~\mathrm{GeV}^2$), with the central line at the midpoints.
  We use bands to emphasize that quasi-diffraction provides a heuristic but not fully precise comparison between hard and forward gapped scattering, as it reflects color-nonsinglet contributions.}
  \label{fig:ratios}
\end{figure*}

\noindent \textbf{\textit{Identifying the two sectors.}}
In practice, to analyze gapped hard scattering in Monte Carlo or experiments, we must choose how we partition a multiparticle state between $p_X$ and $p'$.
Extracting the hadron plane---and hence the azimuth $\phi$ carrying the
discriminating structure functions---requires an operational rule assigning each
particle to a sector. 
It is convenient to choose DIS $1$-jettiness $\tau_1$~\cite{Kang:2012zr,Kang:2013nha,Dotson:2026ttc} as our partition (alternative rules
are reviewed in \S2.3 of \cite{Lee:2025fml}) due to the close relationship between its factorization and Eq.~\eqref{eq:factorization}. We have
\begin{align}\label{eq:tau1-general}
	\tau_1 = \frac{2}{Q^2}\sum_i \min\{\, q_{X_T}\cdot p_i,\ q_{X_C}\cdot p_i\,\}\,,
\end{align}
with light-like reference vectors $q_{X_T}=\tfrac{\omega_{X_T}}{2}\,n_B$ (beam) and
$q_{X_C}=\tfrac{\omega_{X_C}}{2}\,n_J$ (current).
Each particle joins
the sector with the smaller product, partitioning the final state at a boundary rapidity
$\eta_0=\tfrac12\ln(\omega_{X_T}/\omega_{X_C})$,
\begin{align}\label{eq:hemispheres}
	p_X = \!\sum_{i\in \mathcal H_{X_C}}\! p_i\,,
	\qquad
	p' = \!\sum_{i\in \mathcal H_{X_T}}\! p_i\,.
\end{align}
The symmetric choice $\omega_{X_C}=\omega_{X_T}$ is the standard hemisphere
partition in the Breit frame. Note that $\tau_1$ parallels the role of $\beta$ in the diffractive factorization Eq.~\eqref{eq:factorization}, as explained in the Supplemental Material.

The vanishing of $F_{L,3,4}^{\rm gap}$ in Eq.~\eqref{eq:vanishing} is independent of our choice of sector partition in Eq.~\eqref{eq:hemispheres}. 
$H^{\mu\nu}=-g_\perp^{\mu\nu} H(Q^2,\mu)$ is fixed by the collinear directions $n_B$, $n_J$, not by the boundary; thus, it
remains $X$--$Y$ isotropic for any partition.
At leading power, the boundary enters
\textit{only} the soft function $S$ in Eq.~\eqref{eq:factorization}: the collinear sectors sit at large rapidity, far from any finite
$\eta_0$, so displacing the boundary reassigns only wide-angle soft radiation, and the
soft function reduces to the DIS hemisphere soft function with rescaled arguments. In the gapped dijet limit, the reference directions align
with the Breit-frame light-cone axes, $n_J\to\bar n$ (current) and $n_B\to n$ (beam)
with $n_J\cdot n_B\to2$, so that $\mathcal H_{X_C}$ and $\mathcal H_{X_T}$ become the
$\nbar$- and $n$-collinear sectors of the factorization. 

At leading power, Eq.~\eqref{eq:factorization} holds to all orders in $\alpha_s$:
higher-order $\alpha_s$ corrections modify only the scalar 
functions $H$, $\mathcal{B}$, $J$, $S$,
and never generate an
$X$--$Y$ anisotropy, so $F_{L,3,4}^{\rm gap}=0$ holds for the resummed leading-power
result at any order. A nonzero $F_{L,3,4}^{\rm gap}$ in gapped hard scattering can therefore arise
only beyond leading power---from subleading-power corrections carrying the anisotropic structures
absent from $g_\perp^{\mu\nu}$, or from $\cO(\alpha_s)$ real radiation outside the
two-jet configuration entering through fixed-order matching~\cite{Georgi:1977tv,
Cahn:1978se, Cleymans:1978tj, Kopp:1978vx, Mendez:1978zx, Konig:1982uk, Chay:1991nh}. 
For example, spin correlations due to gluon splitting in an additional gluon jet could induce azimuthal dependence.
At $\cO(\alpha_s)$ such modulations are known to scale as powers of $p'_\perp/Q\sim\lambda_t$~\cite{Mendez:1978zx,Bacchetta:2008xw}, and we find below, numerically, that the total effect of
these $\cO(\alpha_s)$ corrections does not notably diminish our ability to discriminate between hard and diffractive events; see Fig.~\ref{fig:ratios}.
Any surviving azimuthal dependence is genuinely a diffractive signature.
In particular, the gapped-hard modulations $F_3^{\rm gap}$ ($\cos2\phi$) and
$F_4^{\rm gap}$ ($\cos\phi$) of Eq.~\eqref{eq:coefficients} switch off as $\beta\to1$,
as we show with RAPGAP below.
\\[-8pt]

\noindent \textbf{\textit{Simulations.}}
We examine our gapped hard scattering analysis with the Monte Carlo generator RAPGAP~\cite{Jung:1993gf}, 
fitting to the unpolarized azimuthal decomposition
\begin{align}\label{eq:fit}
    d\sigma/d\phi = a_{2} + a_4\cos\phi + a_3\cos2\phi\,.
\end{align} 
Comparing Eqs.~\eqref{eq:fit} and \eqref{eq:coefficients}, we have $a_{3,4} \propto F_{3,4}^{\rm gap}$ and  $a_2 \propto (1\!-\!y\!+\!\tfrac{y^2}{2})F_2^{\rm gap}-\tfrac{y^2}{2}F_L^{\rm gap}$. We use the combination $a_2$ because the single $\sqrt s$ value we use cannot separate $F_2^{\rm gap}$ and $F_L^{\rm gap}$.

We generate gapped hard scattering in RAPGAP with the full
$\cO(\alpha_s)$ DIS matrix elements. This is
essential: the discriminating modulations are genuine $\cO(\alpha_s)$
matrix-element effects. 
For every event, we split the final state into the $p'$ and $p_X$ hemispheres through the
$\tau_1$ partition with the Breit hemisphere choice $\eta_0=0$ in Eq.~\eqref{eq:tau1-general}, and form the invariants of Eq.~\eqref{eq:invariants}
directly from the resulting sector momenta.

Figure~\ref{fig:ratios} is our central simulation result. We work at the HERA energy
$\sqrt s=319$~GeV in the narrow window $x\in(0.02,0.04)$, $Q^2\in(800,1000)~\mathrm{GeV}^2$, so that
$y=Q^2/(xs)$ is effectively fixed. 
We bin in $(\beta,\lambda_t)$ and fit the $\phi$ distribution separately in each of the
45 bins. The fits are good: the $\chi^2$ per degree of freedom has a median of 0.96 and
lies between 0.8 and 1.3 in 68\% of the bins.

We plot ratios $a_{3,4}/a_2$ to facilitate comparison of gapped hard scattering to diffraction, for which only structure function ratios $F_i^{\rm quasi}/F_2^{\rm quasi}$ are currently available.
(Quasi-diffraction refers to the sizable subset of forward gapped processes mediated by color-nonsinglet exchange \cite{Lee:2025fml}.)
We check that leading-order or shower-only generators
(e.g.\ Pythia~8~\cite{Bierlich:2022pfr}) give $F_4^{\rm gap}$ consistent with zero
even where they reproduce the $\beta$ and $\lambda_t$ distributions qualitatively, so the
modulation reflects the azimuthal interference of the $\cO(\alpha_s)$ amplitude
rather than the bulk kinematics, consistent with the $\cos\phi$ modulations seen
in inclusive DIS at HERA~\cite{ZEUS:2000esx, ZEUS:2002ohg}. 

Both hard modulations $a_{3,4}/a_2$ fall to zero as $\beta\to1$, realizing the leading-power result
$F_3^{\rm gap}=F_4^{\rm gap}=0$: the azimuthal signal switches off precisely where
the rapidity gap is largest. The bands show the quasi-diffractive prediction of
Ref.~\cite{Lee:2025fml}, which instead stays nonzero and grows toward $\beta\to1$.
However, the comparison is not one-to-one---Ref.~\cite{Lee:2025fml} predicts these ratios
only for the color-nonsinglet quasi-diffractive background, not including
color-singlet diffraction. Our conclusion is independent of this fact; the leading-power
result is that hard scattering supplies \emph{no} $F_L^{\rm gap},F_3^{\rm gap},
F_4^{\rm gap}$ as $\beta\to1$, so \emph{any} azimuthal or longitudinal signal
surviving in the large-gap region is necessarily of diffractive origin---the hard
background is absent, not merely subdominant.
\\[-8pt]

\noindent \textbf{\textit{Conclusions and outlook.}}
We factorize rapidity-gapped hard scattering in DIS, and show that at leading power its hard tensor is purely transverse,
$H^{\mu\nu}\propto g_\perp^{\mu\nu}\propto w_2^{\mu\nu}$. As a consequence the
longitudinal and azimuthal structure functions vanish exactly,
$F_L^{\rm gap}=F_3^{\rm gap}=F_4^{\rm gap}=0$, independently of the sector partition and
to all orders in $\alpha_s$ at leading power. Any azimuthal or longitudinal modulation
surviving in the large-gap region is therefore not a hard-scattering background but a
genuine diffractive signature. RAPGAP simulations are consistent with hard modulations
switching off as $\beta\to1$, realizing this result. This provides evidence that these structure functions are a clean,
model-independent handle for separating diffraction from its hard background at the EIC.
Natural next steps include extending the analysis to subleading power---where the first nonzero $F_{L,3,4}^{\rm gap}$
appear, quantifying the residual hard contamination---to the polarized and nuclear
cases relevant for the EIC program, and to the LHC case \cite{Aretz:2026sjb}.
\\[-8 pt] 

\noindent \textbf{\textit{Acknowledgments.}} 
We thank I. Stewart for helpful discussions. 
S.T.S. appreciates the hospitality of Harvard, MIT, and Washington University in St. Louis during work on this manuscript.
This work was supported by the U.S.~Department of Energy, Office of Science, through the Office of Nuclear Physics, and the LDRD Program at Los Alamos National Laboratory under projects 20230857PRD2, 20240786PRD1 and 20250214ER. LANL is operated by Triad National Security, LLC, for the National Nuclear Security Administration of the U.S.~Department of Energy under Contract No.~89233218CNA000001.

\twocolumngrid
\bibliography{hardly-diffract.bib}
\clearpage

\setcounter{equation}{0}
\renewcommand{\theequation}{S\arabic{equation}}   

\onecolumngrid
\begin{center}
{\large\bf Supplemental Material:\\[2pt]
Minding the gap between hard and diffractive structure functions at the EIC}
\end{center}
\vspace{-1em}
\begin{center}
\small\textit{Equations labeled \eqref{eq:beta-prime}–\eqref{eq:measurement_functions} belong to this Supplemental Material; all other equation numbers, and all citations, refer to the Letter.
Refs.~\cite{Kang:2013nha,Becher:2006qw,Jain:2011iu,Kelley:2011ng, Monni:2011gb, Hornig:2011iu,Manohar:2003vb, Idilbi:2006dg, Becher:2006mr,Dotson:2026ttc} cited here appear in the Letter's bibliography.}
\end{center}
\vspace{-2em}
\subsection{Full factorization formula for gapped hard scattering in DIS}
 
In this Supplemental Material, we present a detailed derivation of the factorization formula Eq.~\eqref{eq:factorization} that
underlies the main text.
We follow the standard $1$-jettiness factorization of
Ref.~\cite{Kang:2013nha}, extended to be fully differential in the diffractive
invariants of Eq.~\eqref{eq:invariants}. Throughout we trade $\beta$ for
\begin{align}
\label{eq:beta-prime}
  \beta'\equiv\frac{1-\beta}{\beta}\,,
\end{align}
which vanishes in the dijet limit together with $t$ and $m_T^2$; the
factorization is most transparent in these variables.
 
It is convenient to expose the invariant measurements as explicit delta
functions inside the hadronic tensor, so we write the cross section as
\begin{align}
\label{eq:final-xsec-meas}
\frac{d^6\sigma}{dx\,dQ^2\,d\beta'\,dt\,dm_T^2\,d\phi}
=\frac{\pi\alpha^2\,y^2}{Q^6}\,
L_{\mu\nu}(x,Q^2,\phi)\,\mathcal{W}^{\mu\nu}_{\rm gap}(x,Q^2,\beta',t,m_T^2,\phi)\,,
\end{align}
where, relative to Eq.~\eqref{eq:cross-section}, the $1/\beta^2$ is cancelled by
rewriting the differential $d\beta$ as $d\beta'$, and the remaining Jacobian
$x/4$ from the change of variables $p'\to(\beta',t,m_T^2,\phi)$ is absorbed into
the measured hadronic tensor $\mathcal{W}^{\mu\nu}_{\rm gap}$. Note that this calligraphic
$\mathcal{W}^{\mu\nu}_{\rm gap}$ is distinct from the main-text $W^{\mu\nu}_{\rm gap}$:
it carries the absorbed Jacobian and the additional measurement deltas, and reads
\begin{align}
\label{eq:W-measured}
  \mathcal{W}^{\mu\nu}_{\rm gap}(x,Q^2,\beta',t,m_T^2,\phi)
  &=\frac{1}{2\pi}\sum_X \widehat{\mathcal M}\;
  (2\pi)^4\delta^4(q+\tau-p_X)\,
  \langle p|J^{\dagger\mu}|X_CX_T\rangle\langle X_CX_T|J^\nu|p\rangle\,.
\end{align}
Here the state sum and phase-space integral are
\begin{align}
\sum_X\equiv\!\!\sum_{X_C,X_T}\!\!
  \bigg[\prod_{i\in X_C}\!\int\!\frac{d^3p_i}{(2\pi)^3 2E_i}\bigg]
  \bigg[\prod_{j\in X_T}\!\int\!\frac{d^3p_j}{(2\pi)^3 2E_j}\bigg],
\end{align}
with $p_X\equiv\sum_{i\in X_C}p_i$ and $p'\equiv\sum_{j\in X_T}p_j$ the total
momenta of the central and target regions, and $\tau=p-p'$. The gap
measurement function $\widehat{\mathcal M}\equiv\delta_{\beta'}\delta_t\delta_{m_T^2}$
collects the three invariant constraints,
\begin{align}
\label{eq:meas-deltas}
  \delta_{\beta'}=\delta\!\Big(\beta'-\tfrac{2q\cdot\tau}{Q^2}+1\Big),\quad
  \delta_{t}=\delta\!\big(t-\tau^2\big),\quad
  \delta_{m_T^2}=\delta\!\big(m_T^2-(p-\tau)^2\big).
\end{align}
The relative azimuthal angle $\phi$ only fixes the orientation of the hadron
plane with respect to the lepton plane, and so does not enter the dynamics; we
return to it at the end.
 
To expose the operator structure, we promote the invariants to measurement
operators built from the total-momentum operators of the two sectors, defined
by their action on a final state $|X_CX_T\rangle$,
\begin{align}
\label{eq:momentum-operators}
  \hat P_{X_C}|X_CX_T\rangle=\Big(\!\sum_{i\in X_C}p_i\Big)|X_CX_T\rangle,
  \quad
  \hat P_{X_T}|X_CX_T\rangle=\Big(\!\sum_{j\in X_T}p_j\Big)|X_CX_T\rangle,
  \quad
  \hat P_X\equiv\hat P_{X_C}+\hat P_{X_T}\,.
\end{align}
With $\hat\tau\equiv p-\hat P_{X_T}$, the measurement function then acts on the
cut through the operators
\begin{align}
\label{eq:measurement-operators}
  \hat t=\hat\tau^2\,,\qquad
  \hat m_T^2=\hat P_{X_T}^2\,,\qquad
  \hat\beta'=\frac{2q\cdot\hat\tau}{Q^2}-1=\frac{\hat m_X^2-\hat t}{Q^2}\,,
  \quad \hat m_X^2=\hat P_{X_C}^2\,,
\end{align}
each returning the corresponding invariant on $|X_CX_T\rangle$. Using
$p_X=\hat P_{X_C}$ and $\tau=p-\hat P_{X_T}$, momentum conservation becomes
$\delta^4(q+\tau-p_X)=\delta^4(q+p-\hat P_X)$. Writing
$(2\pi)^4\delta^4(q+p-\hat P_X)=\int d^4w\,e^{i(q+p-\hat P_X)\cdot w}$,
translating $J^{\dagger\mu}(0)\to J^{\dagger\mu}(w)$, and using completeness of
$|X_CX_T\rangle$ to collapse the state sum, the hadronic tensor
[Eq.~\eqref{eq:W-measured}] reduces to a single operator matrix element on the
proton,
\begin{align}
\label{eq:W-operator}
  \mathcal{W}^{\mu\nu}_{\rm gap}(x, Q^2, \beta',t,m_T^2,\phi)=
  \frac{1}{2\pi}\int d^4w\,e^{iq\cdot w}\,
  \langle p|\,J^{\dagger\mu}(w)\,\widehat{\mathcal M}(\beta',t,m_T^2)\,J^\nu(0)\,|p\rangle.
\end{align}
 
We now match the QCD vector current onto SCET,
\begin{align}
\label{eq:current-matching}
  J^\mu(w)=\sum_f e_f\sum_{n_1 n_2}\sum_{\tilde p_1,\tilde p_2}
  e^{i(\tilde p_1-\tilde p_2)\cdot w}\,
  C_{\alpha\beta}^{\mu}(\tilde p_1,\tilde p_2)\,\bar\chi_{n_1,\tilde p_1}^{(0)\alpha j}(w)\,
  T\big[Y_{n_1}^\dagger Y_{n_2}\big]^{jk}(w)\,
  \chi_{n_2,\tilde p_2}^{(0)\beta k}(w)\,,
\end{align}
where $C_{\alpha\beta}^{\mu}$ is the hard matching coefficient, the soft
interactions have been decoupled by the BPS field redefinition, and the sums run
over the collinear directions $n_{1,2}$ and the label momenta $\tilde p_{1,2}$.
Inserting Eq.~\eqref{eq:current-matching} into Eq.~\eqref{eq:W-operator} and
factorizing the proton state onto a product of independent Hilbert spaces,
\begin{align}
\label{eq:state-factorization}
|p\rangle=|p_{n_B}\rangle\otimes|0_{n_J}\rangle\otimes|0_{us}\rangle\,,
\end{align}
the hadronic tensor separates into collinear, soft, and hard pieces,
\begin{align}
\label{eq:fac-hadron}
  \mathcal{W}^{\mu\nu}_{\rm gap}
  &=\frac{4}{n_J \cdot n_B}\,(2\pi)^3\sum_f e_f^2
  \int dm_X^2\;\delta\Big(\beta'-\frac{m_X^2-t}{Q^2}\Big)
  \int d^2\tilde p_\perp\int dk_{X_C}\,dk_{X_T}\;
  \big[\overline{C}_{\beta\alpha}^{\mu}C_{\alpha'\beta'}^{\nu}\big](\tilde p_1,\tilde p_2)
  \nn\\
  &\times
  \langle 0_{us}|\,[Y_{n_B}^\dagger Y_{n_J}]^{kj}(0)\,
  \delta(k_{X_C}-n_J \cdot \hat p^{\,s}_{X_C})\,\delta(k_{X_T}-n_B \cdot \hat p^{\,s}_{X_T})\,
  [Y_{n_J}^\dagger Y_{n_B}]^{j'k'}(0)\,|0_{us}\rangle
  \nn\\
  &\times
  \Big\{
  \langle p_{n_B}|\,\bar\chi_{n_B}^{\beta k}(0)\,
  \delta\big[m_T^2-\hat m_T^2\big]\,\delta(t-\hat t)\,
  \big[\delta(\bar n_B \cdot q+\bar n_B \cdot \mathcal P)\,
  \delta^2(\tilde p_\perp-\mathcal P_\perp)\,\chi_{n_B}^{\beta' k'}(0)\big]\,|p_{n_B}\rangle
  \nn\\
  &\quad\times
  \langle 0_{n_J}|\,\chi_{n_J}^{\alpha j}(0)\,
  \delta\big[m_X^2-\hat m_X^2\big]\,
  \delta(\bar n_J \cdot q+\bar n_J \cdot \mathcal P)\,
  \delta^2(q_\perp+\tilde p_\perp+\mathcal P_\perp)\,
  \bar\chi_{n_J}^{\alpha' j'}(0)\,|0_{n_J}\rangle
  +(q\to\bar q)
  \Big\}\,,
\end{align}
where label-momentum conservation from the $\int d^4w$ (position space) integration fixes $\tilde p_1=\bar n_J \cdot q\,\frac{n_J}{2}+q_\perp+\tilde p_\perp$ and
$\tilde p_2=-\bar n_B \cdot q\,\frac{n_B}{2}+\tilde p_\perp$ (with $\tilde p_\perp$ the shared transverse label integrated above), $\mathcal P$ is
the label-momentum operator that reads off the label momenta of the SCET fields
and the collinear states, and $\bar{n}_{J,B}^\mu\equiv 2n_{B,J}^\mu/(n_J\cdot n_B)$ are the normalized conjugate vectors to $n_{J,B}^\mu$ satisfying $n_J\cdot \bar{n}_J=n_B\cdot \bar{n}_B=2$. 
 
The $n_J$- and $n_B$-collinear matrix elements, together with their
measurements, define the jet \cite{Becher:2006qw} and beam \cite{Jain:2011iu} functions,
\begin{align}
  &\langle 0_{n_J}|\,\chi_{n_J}^{\alpha j}(0)\,
  \delta\Big[m_X^2-\bar n_J \cdot \mathcal P\big(n_J \cdot \hat p^{\,n_J}+k_{X_C}\big)
    -\mathcal P_\perp^2\Big]\,
  \delta(\bar n_J \cdot q+\bar n_J \cdot \mathcal P)\,
  \delta^2(q_\perp+\tilde p_\perp+\mathcal P_\perp)\,
  \bar\chi_{n_J}^{\alpha' j'}(0)\,|0_{n_J}\rangle
  \nn\\
  &\quad=
  \frac{1}{(2\pi)^3}\frac{\slashed n_J^{\alpha\alpha'}}{4}\,\delta^{jj'}\,
  J_q\big(m_X^2-(\bar n_J \cdot q)\,k_{X_C},\,\mu\big)\,,
\nonumber \\
  &\langle p_{n_B}|\,\bar\chi_{n_B}^{\beta k}(0)\,
  \delta\Big[m_T^2-\bar n_B \cdot \mathcal P\big(n_B \cdot \hat p^{\,n_B}+k_{X_T}\big)
    -\mathcal P_\perp^2\Big]\,
  \delta(t-\hat t)\,
  \big[\delta(\bar n_B \cdot q+\bar n_B \cdot \mathcal P)\,
  \delta^2(\tilde p_\perp-\mathcal P_\perp)\,\chi_{n_B}^{\beta' k'}(0)\big]\,|p_{n_B}\rangle
  \nn\\
  &\quad=r\,\frac{\slashed n_B^{\beta'\beta}}{4}\,\frac{\delta^{kk'}}{N_c}\,
  \mathcal B_q\Big(r\big(m_T^2-\tilde p_\perp^2\big)-(-\bar n_B \cdot q)\,k_{X_T},\,
  -\frac{\bar{n}_B\cdot q}{\bar{n}_B\cdot p},\,\tilde p_\perp^2,\,\mu\Big)\,
  \delta\Big[t + r\,m_T^2-(1+r)\,\tilde p_\perp^2\Big]\,,
\end{align}
where the ratio $r$ appearing in the beam function is
\begin{align}
\label{eq:ratio-r}
  r\equiv\frac{-\bar n_B\cdot q}{\bar n_B\cdot(p+q)}>0\,.
\end{align}
Correspondingly, moving the color and Dirac structures into the soft and hard
pieces gives the soft \cite{Kelley:2011ng, Monni:2011gb, Hornig:2011iu} and hard \cite{Manohar:2003vb, Idilbi:2006dg, Becher:2006mr} functions,
\begin{align}
\label{eq:soft-def}
  S(k_{X_C},k_{X_T},\mu)
  &=\frac{1}{N_c}\,\mathrm{Tr}\,\big\langle 0_{us}\big|\,
  \big[Y_{n_B}^\dagger Y_{n_J}\big](0)\,
  \delta(k_{X_C}-n_J \cdot \hat p^{\,s}_{X_C})\,
  \delta(k_{X_T}-n_B \cdot \hat p^{\,s}_{X_T})\,
  \big[Y_{n_J}^\dagger Y_{n_B}\big](0)\,\big|0_{us}\big\rangle\,,
\nonumber \\
  H_{q\bar q}^{\mu\nu}(q^2,n_J,n_B)
  &=\mathrm{Tr}\!\left[\overline C^{\mu}(q^2,\mu)\,
  \frac{\slashed n_J}{4}\,C^{\nu}(q^2,\mu)\,
  \frac{\slashed n_B}{4}\right]
  =
  -\frac{n_J\cdot n_B}{4}
H(Q^2,\mu)g_{\perp}^{\mu\nu},
\end{align}
where $g_{\perp}^{\mu\nu}$ is the transverse tensor of Eq.~\eqref{eq:hard-tensor}.
As anticipated in the main text, the hard function is purely transverse,
$H_{q\bar q}^{\mu\nu}\propto g_{\perp}^{\mu\nu}$, which is the origin of the
vanishing structure functions $F_{L,3,4}^{\rm gap}=0$.
 
The soft function depends on the soft momenta assigned to $X_C$ and $X_T$, and is
therefore sensitive to the boundary between the two sectors. In general the soft
radiation is partitioned by two reference vectors $q_{X_C}=\omega_{X_C}n_J/2$ and
$q_{X_T}=\omega_{X_T}n_B/2$ through the $1$-jettiness
assignment~\cite{Kang:2013nha,Dotson:2026ttc},
\begin{align}
\label{eq:original-1-jettiness}
  \tau_1=\frac{2}{Q^2}\sum_{i\in X}\min\big\{q_{X_C} \cdot k_i,\;q_{X_T} \cdot k_i\big\}\,,
\end{align}
each particle joining the region ($\mathcal H_{X_C}$ or $\mathcal H_{X_T}$) whose
reference vector gives the smaller product; the boundary may lie at an arbitrary
rapidity. The soft function then reads
\begin{align}
\label{eq:soft-partition}
  S(k_{X_C},k_{X_T},\mu)
  &=\frac{1}{N_c}\,\mathrm{Tr}\sum_{X_s}
  \big|\langle X_s|[Y_{n_J}^\dagger Y_{n_B}](0)|0\rangle\big|^2
  \nn\\
  &\times
  \delta\Big(k_{X_C}-\!\!\sum_{i\in X_s}\!\theta(q_{X_T} \cdot k_i-q_{X_C} \cdot k_i)\,
  n_J \cdot k_i\Big)\,
  \delta\Big(k_{X_T}-\!\!\sum_{i\in X_s}\!\theta(q_{X_C} \cdot k_i-q_{X_T} \cdot k_i)\,
  n_B \cdot k_i\Big)\,.
\end{align}
Because the soft Wilson lines are invariant under a rescaling of their direction
($Y_{\kappa n}=Y_n$ for any $\kappa>0$, boost invariance), we may rescale
the reference vectors to a symmetric configuration.
Defining
\begin{align}
\label{eq:RJB-def}
  R_{X_C}=\sqrt{\frac{\omega_{X_T}\,n_J \cdot n_B}{2\,\omega_{X_C}}}\,,
  \qquad
  R_{X_T}=\sqrt{\frac{\omega_{X_C}\,n_J \cdot n_B}{2\,\omega_{X_T}}}\,,
\end{align}
and $n_J'=n_J/R_{X_C}$, $n_B'=n_B/R_{X_T}$, the partition boundary
$q_{X_T} \cdot k-q_{X_C} \cdot k=\tfrac12\omega_{X_T}R_{X_T}(n_B'-n_J') \cdot k$
becomes symmetric, with $n_J' \cdot n_B'=2$. In terms of the rescaled vectors,
\begin{align}
\label{eq:soft-hemi}
  S(k_{X_C},k_{X_T},\mu)
  =\frac{1}{R_{X_C}R_{X_T}}\,
  S_{\rm hemi}\Big(\frac{k_{X_C}}{R_{X_C}},\frac{k_{X_T}}{R_{X_T}},\mu\Big)\,,
\end{align}
i.e.\ the DIS hemisphere soft function with rescaled momentum arguments and an
overall factor $1/(R_{X_C}R_{X_T})$. The $\eta_0=0$ hemisphere is the special
case $\omega_{X_C}=\omega_{X_T}$, $n_J \cdot n_B=2$ (back-to-back), for which
$R_{X_C}=R_{X_T}=1$ and $S=S_{\rm hemi}$ directly. More generally, in the
back-to-back limit the rescaling factors are pure boosts, $R_{X_C}=e^{+\eta_0}$
and $R_{X_T}=e^{-\eta_0}$, with $\eta_0=\tfrac12\ln(\omega_{X_T}/\omega_{X_C})$
the rapidity of the partition boundary. Fixing $p_X$ and $\tau=p-\hat P_{X_T}$
at the outset pins the collinear sectors, while the residual freedom in the soft
partition is absorbed into the reference vectors, and hence into $R_{X_C,X_T}$.
 
At leading power $n_J$ and $n_B$ align with the current and beam collinear
directions, $n_B\parallel n$ and $n_J\parallel\nbar$, with $n\cdot\nbar=2$.
In this back-to-back limit, $(\bar n_J \cdot q)=(-\bar n_B \cdot q)=Q$ and $r=x/(1-x)$.
Substituting the beam, jet, soft, and hard functions, introducing the
convolution variables $t_{X_C}$ and $t_{X_T}$, performing the $m_X^2$ integral,
and rescaling $k_{X_{C,T}}\to R_{X_{C,T}}\,k_{X_{C,T}}$---which cancels the
$1/(R_{X_C}R_{X_T})$ prefactor of Eq.~\eqref{eq:soft-hemi} against the measure
and restores the standard arguments of $S_{\rm hemi}$---we obtain the fully
differential factorization,
\begin{align}
\label{eq:fac-hadron-BJSH}
  \mathcal{W}^{\mu\nu}_{\rm gap}
  &=-g_\perp^{\mu\nu}
  H(Q^2,\mu)
  \sum_f e_f^2
  \int \frac{d\mathbf{p}_\perp^2}{2}
  \int dt_{X_C} dt_{X_T}dk_{X_C}dk_{X_T}
  \nonumber \\
  &
  \times
  \widehat{\mathcal M}\,
  S_{\rm hemi}\big(k_{X_C},k_{X_T},\mu\big)
  J_q\big(t_{X_C},\,\mu\big)
  \Big\{
  \mathcal B_q\Big(t_{X_T}-\mathbf{p}_\perp^2,\,
  x,\, \mathbf{p}_\perp^2,\,\mu\Big)\,
  +(q\to\bar q)
  \Big\}\,,
\end{align}
in which the measurements $\widehat{\mathcal M}=\delta_{\beta'}\delta_t\delta_{m_T^2}$
are expressed through the beam and jet virtualities ($t_{X_T}$, $t_{X_C}$), the transverse momentum squared $\mathbf{p}_\perp^2$ and the soft momenta ($k_{X_T}$, $k_{X_C}$) as
\begin{align}
\label{eq:measurement_functions}
\delta_{\beta'}&=\delta\Big[\beta'-\frac{t_{X_C}-t+QR_{X_C}k_{X_C}}{Q^2}\Big],
\nonumber \\
\delta_{t}&=\delta\Big[t+r\,m_T^2+(1+r)\,\mathbf{p}_\perp^2\Big],
\nonumber \\
\delta_{m_T^2}&=\delta\Big[m_T^2 - \frac{t_{X_T}-(1+r)\mathbf{p}_\perp^2+QR_{X_T}k_{X_T}}{r} \Big],
\end{align}
and $\mathbf p_\perp^2\equiv-\tilde p_\perp \cdot \tilde p_\perp>0$ is the
(Euclidean) magnitude of the transverse label momentum appearing in the
reductions above. We keep the azimuthal part of the $d^2\tilde p_\perp$
integration separate: the azimuthal angle of $\tilde p_\perp$ is precisely the
relative angle $\phi$ between the hadron and lepton planes, which already appears
as an external variable in Eq.~\eqref{eq:final-xsec-meas}, so integrating over it
would double-count $\phi$.
 
Equation~\eqref{eq:fac-hadron-BJSH} is the comprehensive form of the schematic
factorization $\mathcal{W}^{\mu\nu}_{\rm gap}=H^{\mu\nu}\otimes\mathcal B\otimes
J\otimes S$ in Eq.~\eqref{eq:factorization}. As a check, setting
$\omega_{X_T}=\omega_{X_C}$ in Eq.~\eqref{eq:original-1-jettiness} gives
$R_{X_C}=R_{X_T}=1$, so the boundary becomes the $\eta_0=0$ plane in the Breit
frame; a short calculation then identifies $\beta'$ with the Breit-frame
$1$-jettiness $\tau_1^b$.
Integrating Eq.~\eqref{eq:fac-hadron-BJSH} over $t$, $m_T^2$, and $\phi$
therefore reproduces the $\tau_1^b$ factorization theorem of
Ref.~\cite{Kang:2013nha}, a strong sanity check of our result.

\end{document}